\documentclass[aps,prd,twocolumn,preprintnumbers,nofootinbib,superscriptaddress,amsmath]{revtex4-2}
\usepackage[english]{babel}
\usepackage{bm}
\usepackage[utf8]{inputenc}
\usepackage[colorinlistoftodos, color=green!40, prependcaption]{todonotes}
\usepackage{amssymb}
\usepackage{graphicx}
\usepackage{natbib}

\usepackage[pdftex, pdftitle={Article}, pdfauthor={Author}]{hyperref} 

\begin{document}
\newcommand{\g}[1]{{\color{red}[G: #1]}}
\newcommand{\rs}[1]{{\color{blue}[R: #1]}}
\title{Cosmography with DESI DR2 and SN data} 

\author{Gabriel Rodrigues}
\email{gabrielrodrigues@on.br}
\author{Rayff de Souza}
\email{rayffsouza@on.br}
\author{Jailson Alcaniz}
\email{alcaniz@on.br}
\affiliation{Observatório Nacional, Rio de Janeiro - RJ, 20921-400, Brasil}
\date{\today}

\begin{abstract}
In this paper, we present a kinematic analysis of the Universe's expansion history using cosmography, with a particular emphasis on the jerk parameter $j_0$, which is equal to one in the standard $\Lambda$CDM scenario. We use distance measurements from DESI DR2, both independently and in combination with current Type Ia supernova (SN) samples, to constrain the cosmographic parameters up to the fourth order without relying on a specific cosmological model. Our results show that for the DESI DR2 data alone, the $\Lambda$CDM prediction ($j_0 = 1$) falls within the 2$\sigma$ confidence region. However, when DESI DR2 is combined with the Union3, Pantheon+, and DESY5 SN datasets, the result obtained is discrepant with the $\Lambda$CDM model at about 3.4$\sigma$, 4.1$\sigma$, and 5.4$\sigma$, respectively. These results are consistent with the conclusions based on dark energy parameterizations reported by the DESI Collaboration, which suggest the presence of a dynamic dark energy component in the universe.

\end{abstract}



\maketitle


\section{Introduction}\label{sec:1}
Recent analyses from the Dark Energy Spectroscopic Instrument Collaboration (DESI) have added significant weight to the case for a dynamical dark energy (DDE) component in the Universe. In particular, the combination of Baryon Acoustic Oscillation (BAO) measurements with Type Ia supernova (SN) data has revealed a statistical preference for DDE over the standard $\Lambda$-cold dark matter ($\Lambda$CDM) model, with a significance ranging from 1.7 to 3.3$\sigma$, depending on the supernova sample considered. This preference becomes even stronger — reaching 4.2$\sigma$ — when Cosmic Microwave Background (CMB) data are included, especially in the combination DESI BAO + CMB + 5-year Dark Energy Survey SN (DESY5)~\cite{DESI:2025zgx,DESI:2025fii}.

These results are obtained assuming parameterizations of the dark energy equation-of-state (EoS), $w(a)$, such as the Chevallier–Polarski–Linder (CPL),  $w(a) = w_0 + w_a(1 - a)$~\cite{Chevallier:2000qy,Linder:2002et} or Barboza-Alcaniz (BA) $w(a) = w_0 + w_a(1 - a)/(a^2 + (1 - a)^2)$~\cite{Barboza:2008rh}, which are not a physical model, as they are not grounded in any underlying physical theory. Moreover, the well-known background degeneracy between $w(a)$ parameterization and quintessence models maps these scenarios into the $w_0-w_a$ plane through the cosmic expansion rate $H(z)$, yielding the same cosmological observables~\cite{Scherrer:2015tra,Wolf:2023uno,Shlivko:2024llw,deSouza:2025rhv}. In particular, the $w_0-w_a$ region currently favored by the data predicts a phantom EoS ($w(a) < -1$) over a period in cosmic evolution, an unphysical behavior in general relativity.

The above argument raises an important question: Can we detect signs of deviation from the $\Lambda$CDM model using the same datasets without assuming any specific cosmological model in advance? In this context, nonparametric reconstruction analyses using techniques such as Gaussian processes~\citep{Dinda:2024ktd} and symbolic regression~\citep{Sousa-Neto:2025gpj} have been conducted. These analyses typically show weaker evidence for DDE; see also~\citep{DESI:2024mwx,DESI:2025fii,fuentes25}. This paper follows a different route and
performs a cosmographic analysis of current DESI BAO
and SN data. As well known, cosmography provides a model-independent way to map the expansion history of the Universe, assuming only that the space-time geometry is described by the Friedman-Lemaitre-Robertson-Walker (FLRW) metric and using Taylor expansions of basic observables. From this expansion, one can obtain kinematic parameters such as the Hubble ($H_0$), deceleration ($q_0$), jerk ($j_0$), and higher-order parameters, which can be directly extracted from observables like BAO and SN. In this sense, cosmographic analyses can complement and model independently verify the DESI results obtained from the parametric and non-parametric methods mentioned earlier.

In this work, we use BAO distance measurements from DESI Data Release II together with Type Ia SN data from the DESY5, Pantheon+ and Union3 catalogs, to constrain the cosmographic  parameters and investigate whether DESI data provide evidence for deviations from the standard $\Lambda$CDM model, which predicts $j_0 = 1$, in a completely model-independent way. We structure this paper as follows: Sec. II presents parameters and investigate whether DESI data provide evidence for deviations from the standard $\Lambda$CDM model. In Sec. III we present the equations used to estimate the distance measurements and discuss the results of the satistical analysis. Finally, we present our conclusions in Se. IV.

\section{Cosmography}\label{sec:2}

We assume a flat, homogeneous and isotropic universe described by the FLRW metric
\begin{equation}
ds^2 = dt^2 - a(t)\left[dr^2 + r^2(d\theta^2 + \sin^2\theta d\phi^2)\right]    
\end{equation}
where $t$ is the cosmic time and $a(t)$ is the dimensionless scale factor normalized to unity
at the present time ($a(t_0) = 1$). To analyze the expansion history in a model-independent, cosmogrpahic framework, we follow \cite{Weinberg:1972kfs} and expand the scale factor in a Taylor series around the present cosmic time $t_0$
\begin{align}
\frac{a(t)}{a(t_0)} &= 1 + H_0(t - t_0) - \frac{q_0}{2}H_0^2(t - t_0)^2 
\\ &+ \frac{j_0}{3!}H_0^3(t - t_0)^3  + \frac{s_0}{4!}H_0^4(t - t_0)^4 +\mathcal{O}[(t - t_0)]^5), \nonumber
\end{align}
where the Hubble, deceleration, jerk, and snap parameters are defined as
\begin{align}\label{cp}
H(t) = \frac{\dot{a}}{a}, \quad q(t) = -\frac{1}{H^2} \frac{\ddot{a}}{a}, \nonumber \\ 
j(t) = \frac{1}{H^3} \frac{\dddot{a}}{a}, \quad s(t) = \frac{1}{H^4} \frac{\ddddot{a}}{a}\;.
\end{align}



As discussed in~\cite{cattoen2007}, the series above converges for low redshifts, ${z}\leq 1$. At higher redshifts — where much of the DESI BAO and some of the SN data lie — the expansion can diverge or become inaccurate. In order to circumvent this problem \cite{cattoen2007} proposed a parameterization of the redshift with a new variable $y$, i.e,
\begin{equation}
    y = \frac{z}{1+z},
\end{equation}
which maps $z \in [0, \infty)$ into $y \in [0, 1)$, increasing the convergence radius and encompassing the entire range of observations.  Another issue concerns the truncation of the Taylor expansion. In principle, including more terms to the series improves the accuracy of the approximation to the true expansion history. However, this introduces a larger number of free parameters, which can reduce the statistical significance of the results. Therefore, it is essential to determine the optimal expansion order that maximizes the statistical significance of the fit for the chosen data and provides a better approximation to the true Hubble parameter $H(z)$. 

\begin{figure*}[t]
\centering
    \includegraphics[width=0.8\linewidth]{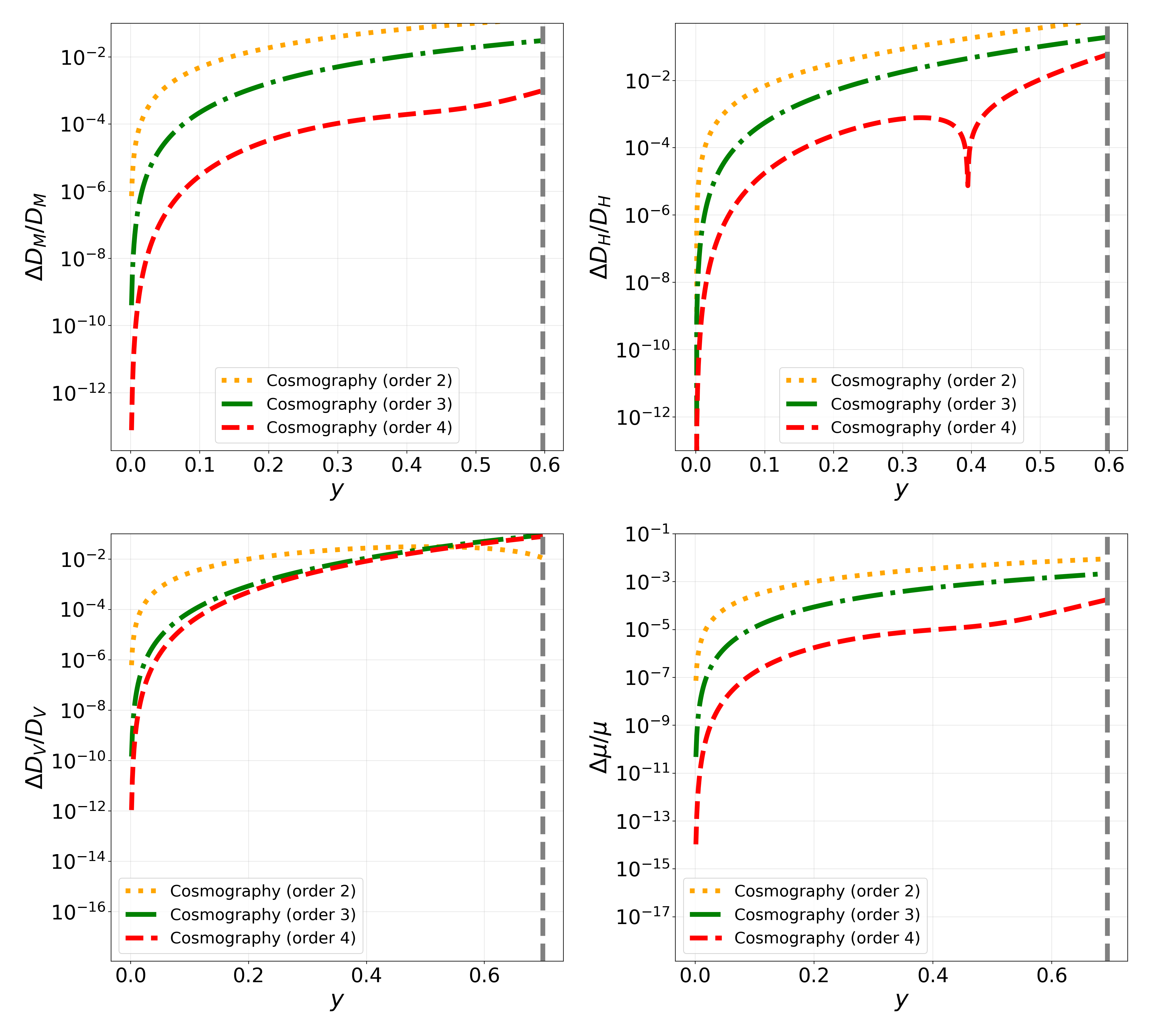}
    \caption{Relative differences between the cosmographic expansion for different orders and the flat $\Lambda$CDM model distance measurements. The dashed gray line indicates the maximum redshift value of the data corresponding to each distance measurement.}
    \label{fig:1b}
\end{figure*}

To determine the appropriate expansion order for our analysis,  we compared the truncated expansions at second, third, and fourth order for the $y$-redshift parameterization with the flat $\Lambda$CDM model assuming $H_0 = 70\; \rm{km/s/Mpc}$ and $\Omega_m = 0.3$.  For this comparison, we fix the parameters $H_0 = 70\; \rm{km/s/Mpc}$, $q_0 = -0.55$, $j_0 = 1$, $s_0 = -0.35$. In Fig.~\ref{fig:1b}, we present the relative difference between the cosmographic expansion and the $\Lambda$CDM distance measurements with the fixed parameters up to the maximum redshift covered by the data. One can observe that even increasing the radius of convergence of the series, the $y$-redshift parameterization remains limited in its ability to accurately reproduce the true evolution of the Hubble parameter $H(y)$ and its associated distance measures. To improve the precision of the cosmographic expansion, it is necessary to consider an optimal combination of the series expansion order and the redshift range being analyzed. 

As shown, the relative differences for the comoving distance $D_M(y)$ and the luminosity distance $D_L(y)$ remain small even for a third-order expansion up to $y \approx 0.7$ (corresponding to $z \approx 2.3$), which encompasses the full range of supernova data. The effective volume distance $D_V(y)$ is well reproduced even at second order, which becomes more accurate than the higher orders starting from $y \approx 0.5$. Nevertheless, all orders of the $D_V(y)$ expansion maintain a relative difference below 10\% up to $y = 0.7$. The most challenging case is the Hubble distance $D_H$, which is directly related to the Hubble parameter $D_H \propto 1/H$. It exhibits significant discrepancies even at relatively small values of $y$.

In this case, the choice of expansion order and the maximum redshift becomes particularly critical. For this reason, the pair of points from the DESI DR2 Ly$\alpha$ forest BAO measurements, $D_H$ and $D_M$ at $z = 2.33$, were converted into the effective volume distance $D_V$. This allowed for a more reliable fourth-order expansion approximation of $D_H$ up to $y \approx 0.6$, with an accuracy of approximately 8\%, while maintaining the relative difference in $D_V$ at a similar level of about 8\% at $y \approx 0.7$. Therefore, throughout this work, our results will be presented using a cosmographic series truncated at the fourth-order expansion, which yields the smallest relative differences for all distance measurements used, while also providing good statistical significance to the results.


\section{Methodology and Results}\label{sex:3}

To constrain the kinematic parameters defined in (\ref{cp}), we perform a Markov Chain Monte Carlo (MCMC) analysis combining measurements of the transverse comoving angular distance ($D_M/r_s$), the Hubble distance ($D_H/r_s$), and the effective volume distance ($D_V/r_s$) from DESI DR2, together with Type Ia SN data from DESY5, Pantheon+, and Union3 compilations. For this purpose, we develop a likelihood module and interfaced with \textit{Cobaya}~\cite{Torrado_2021} and \texttt{emcee}~\cite{Foreman_Mackey_2013} to obtain the MCMC statistical results\footnote{The code is publicly available at \href{https://github.com/Gabriel-Rodrigues1/cosmography_mcmc}{GitHub}.}. The likelihood function used in our analysis assumes Gaussian-distributed errors for both BAO and supernovae data, and can be written as
\begin{equation}
\mathcal{L}(\boldsymbol{\theta}) \propto 
\exp\left[-\tfrac{1}{2}\chi^2(\boldsymbol{\theta})\right],
\qquad 
\chi^2(\boldsymbol{\theta}) = \chi^2_{\rm BAO}(\boldsymbol{\theta}) + \chi^2_{\rm SN}(\boldsymbol{\theta}),
\end{equation}
where $\boldsymbol{\theta} = \{H_0, q_0, j_0, s_0\}$ are the cosmographic parameters. For he DESI BAO measurements we use
\begin{equation}
\chi^2_{\rm BAO}(\boldsymbol{\theta}) = 
\bigl(\mathbf{d}_{\rm BAO} - \mathbf{m}_{\rm BAO}(\boldsymbol{\theta})\bigr)^{\!\rm T}
\, \mathbf{C}_{\rm BAO}^{-1}\,
\bigl(\mathbf{d}_{\rm BAO} - \mathbf{m}_{\rm BAO}(\boldsymbol{\theta})\bigr),
\end{equation}
where $\mathbf{d}_{\rm BAO}$ denotes the observed DESI data vector (comprising $D_M/r_s$, $D_H/r_s$, and $D_V/r_s$ at different redshifts), $\mathbf{m}_{\rm BAO}(\boldsymbol{\theta})$ are the corresponding cosmographic predictions, and $\mathbf{C}_{\rm BAO}$ is the DESI covariance matrix. For the supernovae dataset we similarly have
\begin{equation}
\chi^2_{\rm SN}(\boldsymbol{\theta}) = 
\bigl(\boldsymbol{\mu}_{\rm obs} - \boldsymbol{\mu}_{\rm th}(\boldsymbol{\theta})\bigr)^{\!\rm T}
\, \mathbf{C}_{\rm SN}^{-1}\,
\bigl(\boldsymbol{\mu}_{\rm obs} - \boldsymbol{\mu}_{\rm th}(\boldsymbol{\theta})\bigr),
\end{equation}
where $\boldsymbol{\mu}_{\rm obs}$ are the observed distance moduli, $\boldsymbol{\mu}_{\rm th}(\boldsymbol{\theta})$ are the theoretical predictions from the cosmographic expansion, and $\mathbf{C}_{\rm SN}$ is the  the corresponding covariance matrix (including both statistical and systematic uncertainties). To perform our analyses, we set the priors on the cosmographic parameters as $H_0 \in [50,70]$, $q_0 \in [-2,2]$, $j_0 \in [-5,5]$, and $s_0 \in [-10,10]$. Table~\ref{tab:bao_data} shows the geometric BAO data points from DESI DR2 used in the analysis. 

\begin{table}[t]
\centering
\begin{tabular}{cccc}
\hline
\textbf{z} & \textbf{Type} & \textbf{Value} & \boldmath$\sigma$ \\
\hline
0.29  & $D_V/r_s$ & 7.941  & 0.076 \\
0.51  & $D_M/r_s$ & 13.587 & 0.168 \\
0.51  & $D_H/r_s$ & 21.862 & 0.428 \\
0.70  & $D_M/r_s$ & 17.350 & 0.179 \\
0.70  & $D_H/r_s$ & 19.455 & 0.333 \\
0.93  & $D_M/r_s$ & 21.575 & 0.161 \\
0.93  & $D_H/r_s$ & 17.641 & 0.201 \\
1.32  & $D_M/r_s$ & 27.600 & 0.324 \\
1.32  & $D_H/r_s$ & 14.176 & 0.224 \\
1.48  & $D_M/r_s$ & 30.511 & 0.763 \\
1.48  & $D_H/r_s$ & 12.817 & 0.518 \\
2.33  & $D_V/r_s$ & 31.269 & 0.256 \\
\hline
\end{tabular}
\caption{DESI DR2 BAO observables and their uncertainties.}
\label{tab:bao_data}
\end{table}

\subsection{Cosmographic expressions}

In what follows, we present the cosmographic distance measurements used in our study up to the snap $s_0$ parameter in the $y$-redshift parameterization:

\begin{itemize}

\item Hubble parameter:
\begin{equation}
    H(y) \approx H_0 (H_1+H_2+H_3 + H_4),
\end{equation}
\begin{align}
H_1 &=1  \nonumber \\
H_2 &= (1 + q_0) y   \nonumber\\
H_3 &= \frac{1}{2}\left( 2 + 2q_0 - q_0^2 + j_0  \right) y^2  \nonumber \\
H_4 &= \frac{1}{6}\left( 6 + 3q_0^3 - 3q_0^2 + 6q_0 + 3j_0 - 4j_0q_0 - s_0  \right) y^3  \nonumber
\end{align}

\item Hubble distance: 
\begin{equation}
D_H(y) = \frac{c}{H_0} (D_{H_1}+D_{H_2}+D_{H_3} + +D_{H_4}),
\end{equation}
\begin{align}
 D_{H_1} &= 1  \nonumber \\
 D_{H_2} &= - (1 + q_0) y  \nonumber\\ 
 D_{H_3} &= \left( q_0 + \frac{3}{2}q_0^2 - \frac{1}{2}j_0 \right) y^2  \nonumber \\
 D_{H_4} &= \left( -\frac{5}{2}q_0^3 - \frac{3}{2}q_0^2 + \frac{1}{2}j_0 - \frac{5}{3}q_0j_0 + \frac{1}{6}s_0 \right) y^3  \nonumber
\end{align}

\item Comoving distance,
\begin{equation}
D_M(y) = \frac{c}{H_0}(D_{M_1}+D_{M_2}+D_{M_3} + D_{M_4} ),
\end{equation}
\begin{align}
  D_{M_1} &= y  \nonumber \\
  D_{M_2} &= \frac{1}{2}(1 - q_0)y^2 \nonumber \\
  D_{M_3} &= \frac{1}{6} ( 2 - 2q_0 + 3q_0^2 - j_0 ) y^3    \nonumber \\
  D_{M_4} &= \frac{1}{24} ( 6 - 6q_0 + 9q_0^2 - 15q_0^3 - 3j_0 + 10q_0j_0 + s_0 ) y^4    \nonumber
\end{align}

\item Effective volume distance: 
\begin{equation}
D_V(y) = \frac{c}{H_0} (D_{V_1}+D_{V_2}+D_{V_3} +D_{V_4}),
\end{equation}
\begin{align}
  D_{V_1} &=  y \nonumber \\
  D_{V_2} &= \frac{1}{3}(1 - 2q_0)y^2 \nonumber \\
  D_{V_3} &= \frac{1}{36} \left( 7 - 10q_0 + 29q_0^2 - 10j_0  \right) y^3 \nonumber \\
  D_{V_4} &= \frac{1}{324} ( 44 - 57q_0 + 117q_0^2 - 376q_0^3 - 39j_0 \nonumber
 \\&+ 258q_0j_0 + 27s_0)y^4 \nonumber
\end{align}

\item Distance modulus:
\begin{equation}
\mu(y) = 5 \log_{10} \left( D_L(y) \right) + 25,
\end{equation}
where $D_L(y) = \frac{D_M(y)}{1 - y}$.    
\end{itemize}
The distances mentioned above are measured in megaparsecs (Mpc). In our analysis, we set the value of the sound horizon at $r_s = 148.3\pm 4.3$\;Mpc obtained from cosmic clocks together with SN and BAO data~\cite{Verde_2017}. We also tested different values of $r_s$, which did not result in significant changes to our conclusions.



\begin{figure*}[t]
\centering
\includegraphics[width=0.76\linewidth]{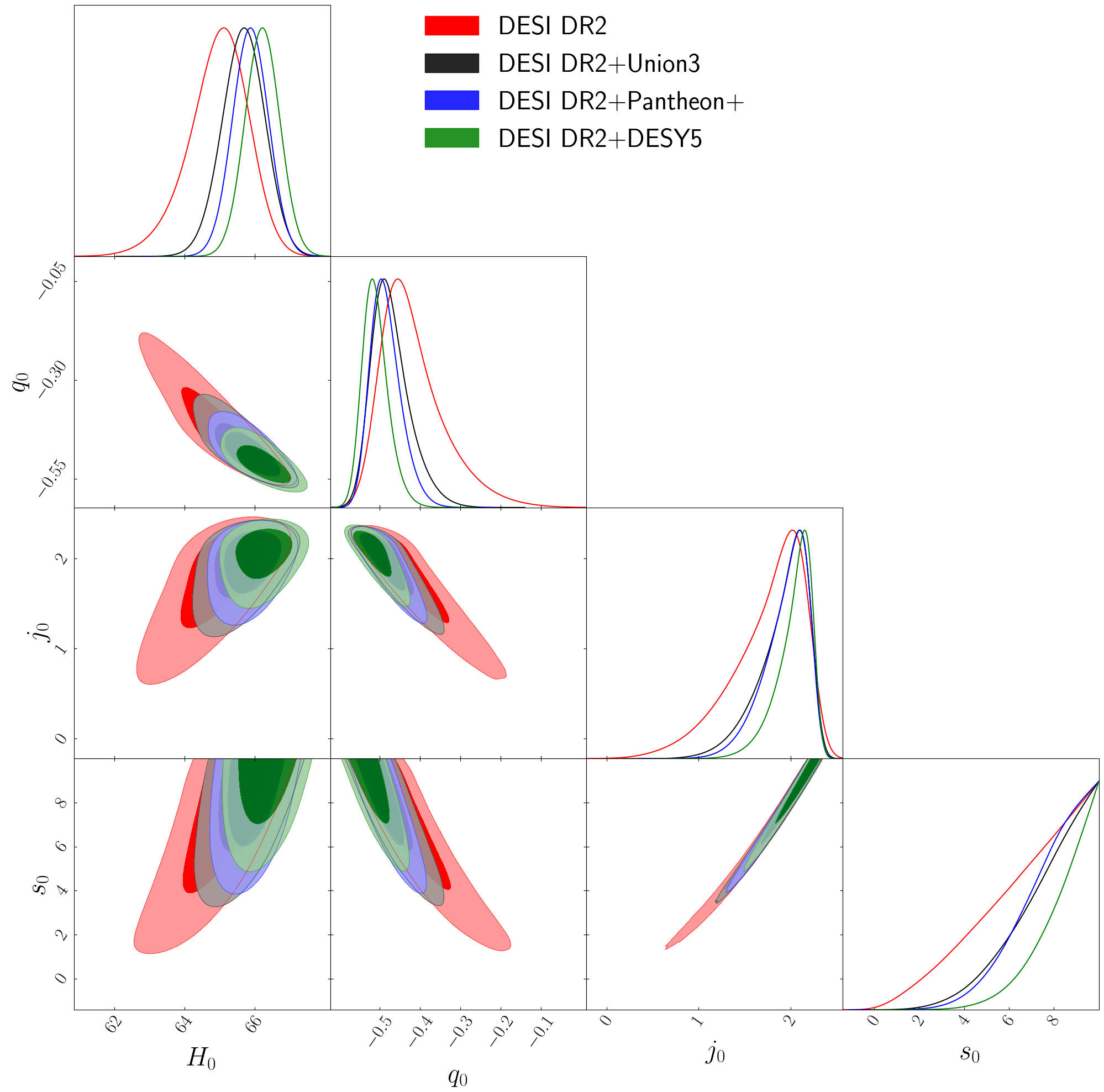}
\caption{Confidence contours for the kinematic parameters at 68\% and 95\% confidence levels obtained with DESI DR2 and current SN samples.}
\label{fig:3}
\end{figure*}

\begin{table*}[]
    \centering
    \begin{tabular}{|l|c|c|c|c|}
        \hline
        \hline
         Parameter & DESI DR2 & DESI DR2 + Union3  & DESI DR2 + Pantheon+  & DESI DR2 + DESY5\\
        \hline
        $j_0$                  & $1.77^{+0.61}_{-0.86}$   & $1.93^{+0.41}_{-0.58}$     &     $1.96^{+0.37}_{-0.50}$       &  $2.04^{+0.30}_{-0.42}  $ \\
        $q_0$                  & $-0.42^{+0.17}_{-0.13} $ & $-0.47^{+0.10}_{-0.086}$   &   $-0.48^{+0.079}_{-0.069} $     &  $-0.51^{+0.068}_{-0.060}  $ \\
        $H_0$                  & $65^{+1.6}_{-1.8}$       & $66^{+1.2}_{-1.2}$         &   $66^{+1.0}_{-1.0}     $        &  $66^{+0.97}_{-0.97}     $ \\
        $s_0$                  & $> 2.63$                 & $> 4.57  $                 &   $> 5.01    $                   &  $   > 5.93  $ \\
        $\chi^2_{dof}$                  & $ 1.94$                 & $ 1.42 $                 &   $  1.04  $                   &  $  1.01 $ \\

        \hline
        \hline
    \end{tabular}
    \caption{Mean and $2\sigma$ limits for the kinematic parameters obtained with DESI DR2 and the SN samples discussed in the text. The reduced chi-square, $\chi^2_{\rm dof}$, is also reported for each dataset to quantify the goodness of fit.} 
    \label{Tab1}
\end{table*}

\subsection{$\Lambda$CDM Model}

The jerk parameter describes how the acceleration of the Universe evolves over time, $j(t) = \frac{1}{H^3} \frac{\dddot{a}}{a}$. The third derivative of $a(t)$ with respect to cosmic time $t$ is given by
\begin{align} 
\dddot{a} &= a(H^3 +3H\dot{H}+\ddot{H})\;.
\end{align}
where $H^2(a) = H_0^2 \left[ \Omega_{m} a^{-3}  + (1 -\Omega_{m}) \right]$ for the flat $\Lambda$CDM model. Substituting the above expressions into $j(t)$, we find
\begin{equation} 
j(t) = \frac{1}{H^3}(H^3 +3H\dot{H}+\ddot{H})\;.
\end{equation}
The first and second derivatives of the Hubble parameter with respect to cosmic time $t$ are,
\begin{align} 
\dot{H} &= -\frac{3H_0^2}{2}\Omega_m a^{-3}\;, \\
\ddot{H} &=\frac{9H_0^2 H}{2}\Omega_m a^{-3}\;.
\end{align}
Substituting the above expression in the $j(t)$ definition, we find $j(t) = 1$. Therefore, similarly to the standard dynamical approach, which is particularly powerful in testing the robustness of the standard model from deviations of $w(a) = -1$, the above result makes the $j(t)$ parameter a useful benchmark whose constraints can also be used to check for departures from the flat $\Lambda$CDM model. In this case, any observed deviation from $j_0 = 1$, could point to the presence of dynamical dark energy or other new physics beyond the standard $\Lambda$CDM paradigm.

\begin{figure}[t]
\centering
\includegraphics[width=0.75\linewidth]{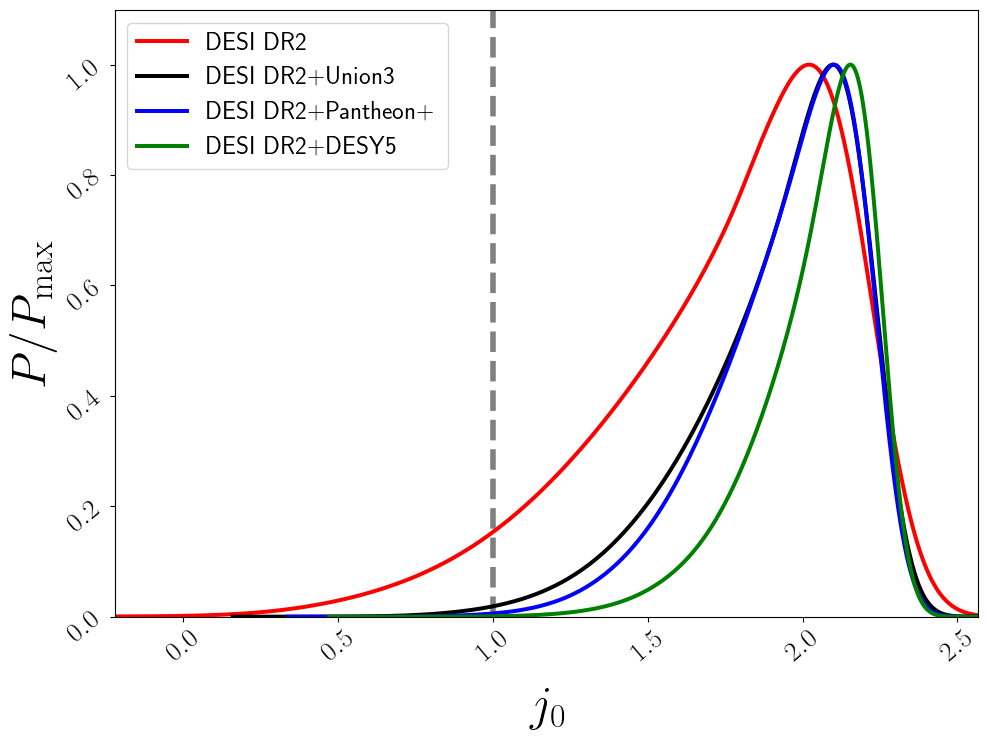} 
\includegraphics[width=0.75\linewidth]{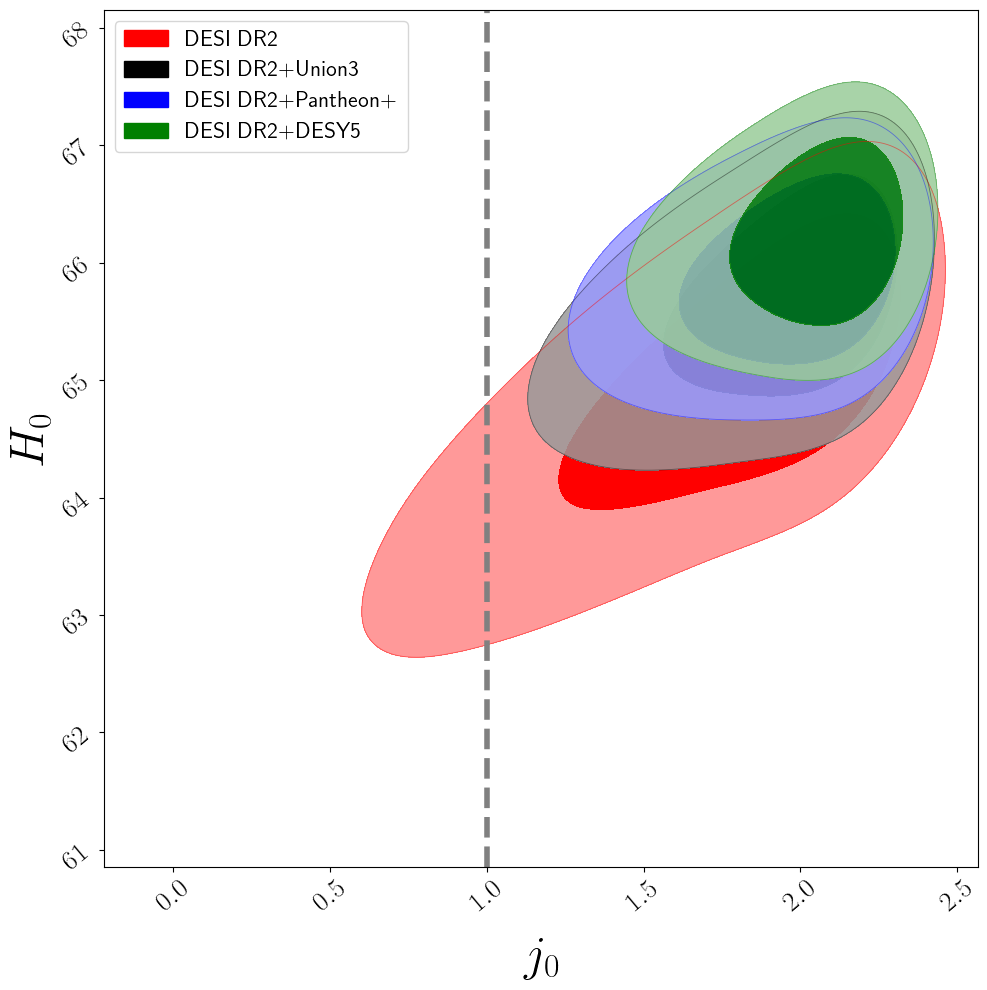}
\caption{Constraints on the kinematic parameters at 68\% and 95\% confidence levels for DESI DR2, with and without the inclusion of different SN samples. The top panel shows the marginalized posterior distribution of $j_0 $ for the datasets considered while the bottom panel presents the confidence contours in the $j_0$–$H_0$ plane.}
\label{fig:2}
\end{figure}

Figures~\ref{fig:2} and \ref{fig:3} present our constraints on the kinematic parameters at $1\sigma$ and 2$\sigma$ confidence level -- see also Table~\ref{Tab1}. The top panel of Fig.~\ref{fig:3} shows the marginalized posterior distribution of $j_0$ obtained from DESI DR2 distance measurements combined with different SN samples while the bottom panel shows the $1\sigma$ and 2$\sigma$ confidence contours in the $j_0$-$H_0$ plane. {For DESI DR2 alone, the reference value $j_0 = 1$ lies within the $2\sigma$ confidence level. However, when DESI DR2 is combined with Union3 or Pantheon+, the value $j_0 = 1$ falls outside the $2\sigma$ range, lying $3.4\sigma$ away in the case of Union3 and $4.1\sigma$ away for Pantheon+. The most significant deviation occurs when DESI DR2 is combined with DESY5, where $j_0$ is found to be $5.4\sigma$ from the reference value.} These results agree with the  analysis reported by the DESI collaboration using $w(a)$ parameterizations, in which the inclusion of SN samples increases the statistical significance of the deviation from the standard $\Lambda$CDM model. 

This deviation from $j_0 = 1$ may suggest a preference for DDE models with a time-evolving equation of state (EoS). In particular, the values of $j_0 > 1$ can be associated with freezing quintessence scenarios, in which the EoS parameter $w(z)$ decreases over time and approaches asymptotically $w=-1$ from above. In these models, the field slows down and freezes as the Universe expands, leading to a dark energy behavior that becomes increasingly similar to a cosmological constant at late times. 

As representative examples, we can take the generic CPL parameterization of DDE. 
In this case, the jerk parameter is obtained following Eqs. (11) - (13) evaluated at $t = t_0$, such that
\begin{equation}
    j_0 = 1 + \frac{3}{2}(1-\Omega_m)\left[ 3 w_0 (1+w_0) + w_a \right]\; .
\end{equation}
The $j_0 > 1$ condition implies that $w_a > -3 w_0 (1+w_0)$. Thus, for the current experimental constraints on $w_0 \sim -0.8$ \cite{DESI:2025zgx}, $w_a$ would have to be positive in order to push the jerk parameter towards the $j_0 > 1$ region. For the BA parameterization, 
the jerk parameter becomes
\begin{equation}
    j_0 = 1+\frac{3}{2}(1-\Omega_m)[3(1+w_0)(w_0+w_a) - w_a (3 w_0 +2)]\; ,
\end{equation}
such that the $j_0 > 1$ condition amounts to $w_a < 3(1+w_0)(w_0+w_a)/(3 w_0 + 2)$, which is also disfavored by data.

Finally, for a parameterization that generalizes a class of quintessence potentials~\cite{Carvalho:2006fy}, $ w(a) = -1 + \alpha a^\beta $, where for $\alpha > 0$ and $\beta > 0$ it reproduces the thawing behavior of scalar field models and for $\alpha > 0$ and $\beta < 0$ the freezing -- for a more comprehensive analysis focusing on the thawing behavior, see~\cite{deSouza:2025rhv,gabriel25}. Assuming the above dark energy EoS, the jerk parameter is given by
\begin{equation}
j_0 = 1 - \frac{3}{2}(1-\Omega_m)\left[ 3\alpha (1+\alpha) + \alpha\beta \right],
\end{equation}
with values of $\alpha > 0$ and $\beta < -3(1+\alpha)$ leading to $j_0 > 1$. As shown in \cite{deSouza:2025rhv}, current data strongly favor high positive values of $\beta$, which points towards the thawing behavior of quintessence. Therefore, the preference for $j_0 > 1$, obtained in the cosmographic approach, seems to go against the current constraints on parametric analyses of DDE.  

\section{Final Remarks}

In this work, we investigated cosmographic constraints from DESI DR2 BAO and SN distance measurements, with a focus on the jerk parameter, which is equal to 1 in  the $\Lambda$CDM model. 
Therefore, any significant deviation from this value provides a powerful, model-independent probe of the standard cosmological framework.

Our analysis reveals that DESI DR2 data alone yields a value of $j_0$ that is marginally consistent with the $\Lambda$CDM prediction at the $2\sigma$ level. However, when DESI DR2 is combined with the DESY5 Supernova dataset, the inferred value of $j_0$ shifts further from unity, indicating an increasing tension with the standard model (see Table~\ref{Tab1}). Notably, the combined DESI DR2 + DESY5 dataset exhibits the most pronounced deviation from $j_0 = 1$ at more than $5\sigma$, marking a significant departure from the $\Lambda$CDM expected value.

These results indicate a clear deviation from the $\Lambda$CDM model, providing evidence against the standard cosmology based on a purely kinematic approach. Additionally, for some data combination, they show disagreement with recent non-parametric analyses of the same data sets~~\citep{Dinda:2024ktd,Sousa-Neto:2025gpj} while aligning with the conclusions drawn from dark energy parameterizations reported by the DESI Collaboration, which suggest the presence of a dynamic dark energy component in the universe~\cite{DESI:2025zgx,DESI:2025fii}. Thus, the results of this study offer a complementary and model-independent consistency check to those obtained under specific cosmological assumptions.


\section*{Acknowledgements}

GR and RdS are supported by the Coordena\c{c}\~ao de Aperfei\c{c}oamento de Pessoal de N\'ivel Superior (CAPES). JA is supported by CNPq grant No. 307683/2022-2 and Funda\c{c}\~ao de Amparo \`a Pesquisa do Estado do Rio de Janeiro (FAPERJ) grant No. 259610 (2021).

\bibliographystyle{apsrev4-2}
\bibliography{references}

\label{lastpage}

\end{document}